\documentclass[conference]{IEEEtran} 
\IEEEoverridecommandlockouts
\usepackage[binary-units,per-mode=symbol,per-symbol=p]{siunitx}
\usepackage{cite}
\usepackage{adjustbox}
\usepackage[ruled,vlined]{algorithm2e}
\usepackage{amsmath,amssymb,amsfonts}
\usepackage{algorithmic}
\usepackage{graphicx}
\usepackage{textcomp}
\usepackage[colorlinks=false, pdfborder={0 0 0}, breaklinks]{hyperref}
\usepackage{xcolor}
\usepackage{balance}
\usepackage{enumitem}
\setlist{nolistsep,leftmargin=*}
\usepackage{setspace}
\usepackage{wrapfig}
\usepackage{listings}
\usepackage{color}
\usepackage{caption}
\usepackage{subcaption}
\usepackage{todonotes}
\usepackage[longtable]{multirow}

\def\BibTeX{{\rm B\kern-.05em{\sc i\kern-.025em b}\kern-.08em
    T\kern-.1667em\lower.7ex\hbox{E}\kern-.125emX}}

\usepackage{amsmath}
\DeclareMathOperator*{\argmax}{arg\,max}

\usepackage{ctable}
\newcommand{\cf}{\emph{cf.}\xspace}

\newcommand{\ie}{\emph{i.e.}, }
\newcommand{\eg}{\emph{e.g.}, }
\newcommand{\etc}{\emph{etc. }}
\newcommand{\HAS}{\emph{HTTP Adaptive Streaming }}

\newcommand{\VVC}{\emph{Versatile Video Coding }}

\newcommand{\DCT}{\emph{discrete cosine transform }}

\newcommand{\HLS}{\emph{HTTP Live Streaming }}

\newcommand{\jnd}{\emph{just noticeable difference}\xspace}
\newcommand{\vsr}{video super-resolution\xspace}
\newcommand{\vig}[1]{\textcolor{black}{#1}}

\newcommand{\EY}{$E_{\text{Y}}$}
\newcommand{\h}{$h$}
\newcommand{\LY}{$L_{\text{Y}}$}

\newcommand{\BDRP}{$BDR_{\text{P}}$}
\newcommand{\BDRV}{$BDR_{\text{V}}$}

\newcommand{\opte}{\texttt{OPTE}\xspace}

\newcommand{\scheme}{\texttt{ViSOR}\xspace}

\newcommand*\circled[1]{\tikz[baseline=(char.base)]{%
            \node[shape=circle,fill=white!20,draw,inner sep=0.5pt] (char) {#1};}}
            
\begin{document}
\bstctlcite{IEEEexample:BSTcontrol}
\title{Video Super-Resolution for Optimized Bitrate and Green Online Streaming}

\author{\IEEEauthorblockN{Vignesh V Menon\IEEEauthorrefmark{1}, Prajit T Rajendran\IEEEauthorrefmark{2}, Amritha Premkumar\IEEEauthorrefmark{3}, Benjamin Bross\IEEEauthorrefmark{1}, Detlev Marpe\IEEEauthorrefmark{1}
}
\IEEEauthorblockA{\IEEEauthorrefmark{1}Video Communication and Applications Department, Fraunhofer HHI, Berlin, Germany}
\IEEEauthorblockA{\IEEEauthorrefmark{2}Universite Paris-Saclay, CEA, List, F-91120, Palaiseau, France}
\IEEEauthorblockA{\IEEEauthorrefmark{3}Department of Computer Science, Rheinland-Pfälzische Technische Universität, Kaiserslautern, Germany}
\vspace{-2.6em}
}

\maketitle

\begin{abstract}
Conventional per-title encoding schemes strive to optimize encoding resolutions to deliver the utmost perceptual quality for each bitrate ladder representation. Nevertheless, maintaining encoding time within an acceptable threshold is equally imperative in online streaming applications. Furthermore, modern client devices are equipped with the capability for fast deep-learning-based \vsr~(VSR) techniques, enhancing the perceptual quality of the decoded bitstream. 
This suggests that opting for lower resolutions in representations during the encoding process can curtail the overall energy consumption without substantially compromising perceptual quality. In this context, this paper introduces a \underline{vi}deo \underline{s}uper-resolution-based latency-aware \underline{o}ptimized bit\underline{r}ate encoding scheme (\scheme) designed for online adaptive streaming applications. \scheme determines the encoding resolution for each target bitrate, ensuring the highest achievable perceptual quality after VSR within the bound of a maximum acceptable latency. 
Random forest-based prediction models are trained to predict the perceptual quality after VSR and the encoding time for each resolution using the spatiotemporal features extracted for each video segment. Experimental results show that \scheme targeting fast super-resolution convolutional neural network (\texttt{FSRCNN}) achieves an overall average bitrate reduction of \SI{24.65}{\percent} and \SI{32.70}{\percent} to maintain the same PSNR and VMAF, compared to the HTTP Live Streaming (HLS) bitrate ladder encoding of 4\,s segments using the x265 encoder, when the maximum acceptable latency for each representation is set as two seconds. Considering a just noticeable difference (JND) of six VMAF points, the average cumulative storage consumption and encoding energy for each segment is reduced by \SI{79.32}{\percent} and \SI{68.21}{\percent}, respectively, contributing towards greener streaming.
\end{abstract}


\section{Introduction}
\HAS (HAS) has become the \textit{de-facto} standard in delivering video content for various clients regarding internet speeds and device types. The main idea behind HAS is to divide the video content into segments and encode each segment at various bitrates and resolutions, called \textit{representations}, which are stored in plain HTTP servers. These representations are stored to continuously adapt the video delivery to the network conditions and device capabilities of the client~\cite{Bentaleb2019}. Traditionally, a fixed bitrate ladder, \eg \HLS (HLS) bitrate ladder~\cite{HLS_ladder_ref}, is used in online streaming applications. However, recently, \textit{per-title encoding}~\cite{netflix_paper} is introduced for online streaming, \ie estimating encoding resolution~\cite{gnostic,res_pred_ref1,faust_ref}, \vig{bitrate}~\cite{vvenc_qp_pred}, framerate~\cite{cvfr_ref}, encoding preset~\cite{nasiri_multi_preset}, CPU resources~\cite{jale_ref}, \etc which yield the highest perceptual quality for a given video content. Notably, in online streaming, where real-time viewing is crucial, encoding delays are typically expected to be very low, often in a few seconds or less. The minimal delay ensures viewers receive the action as close to real-time as possible~\cite{pradeep_ref}. Moreover, reducing encoding time is also essential, resulting in lower energy consumption at the data centers during the encoding process. This is particularly important as environmental consciousness grows and companies seek to adopt greener practices \vig{to reduce energy consumption in the streaming chain, \ie encoding (server-side), transmission (network-side), and decoding (client-side)}. Fig.~\ref{fig:vsr_intro_fig} shows the rate versus encoding time curve of two representative sequences encoded at various encoding resolutions. It is observed that lowering encoding resolution can contribute to lowering the encoding latency.

\subsubsection*{Video super-resolution for green streaming} 
\begin{figure}[t]
\centering
\begin{subfigure}{0.95\columnwidth}
\centering
\includegraphics[width=0.47\columnwidth]{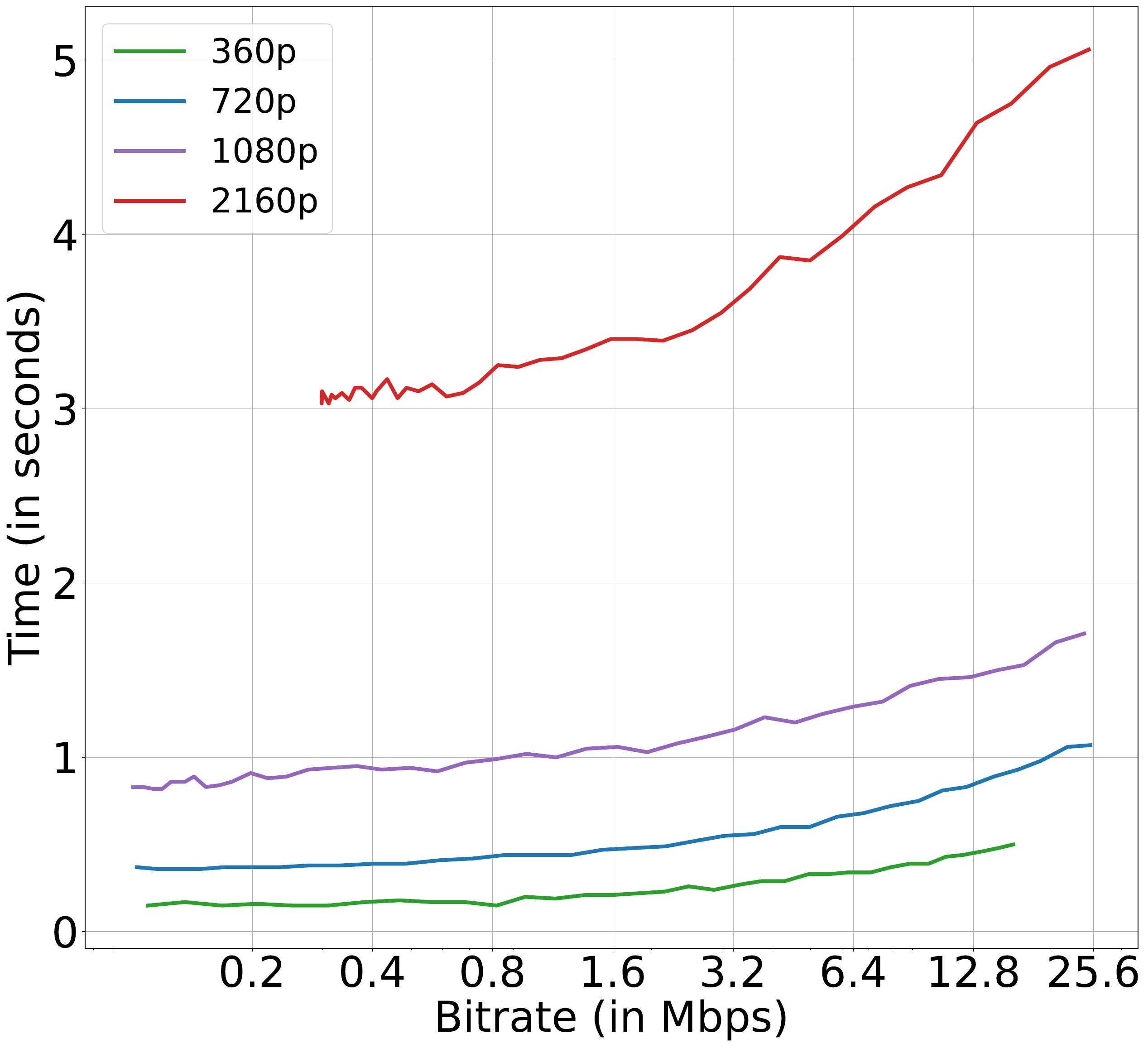}
\includegraphics[width=0.495\columnwidth]{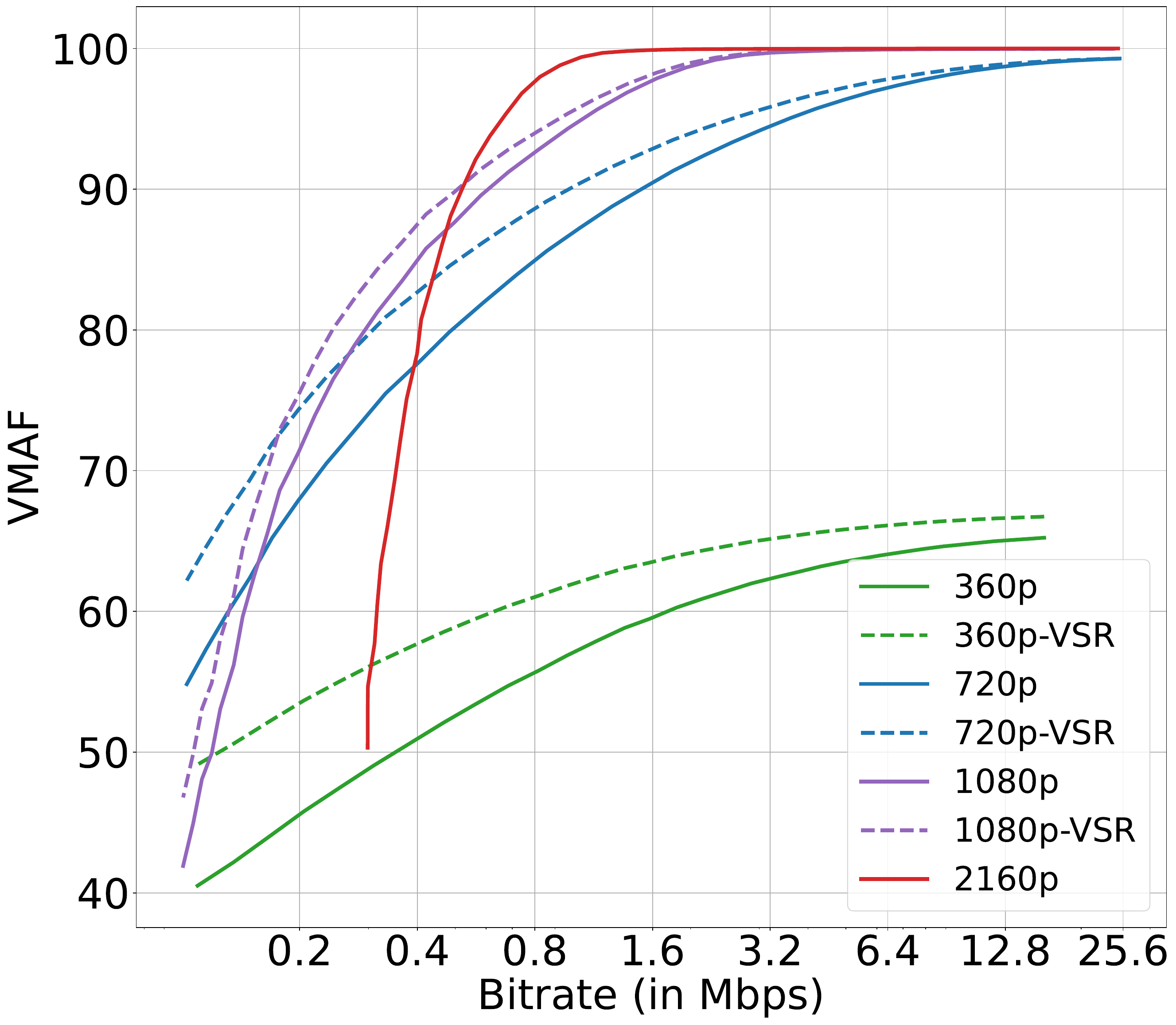}
\vspace{-0.2em}
\caption{\textit{Characters\_s000}}
\label{fig:intro_char}
\end{subfigure}
\begin{subfigure}{0.95\columnwidth}
\centering
\includegraphics[width=0.47\columnwidth]{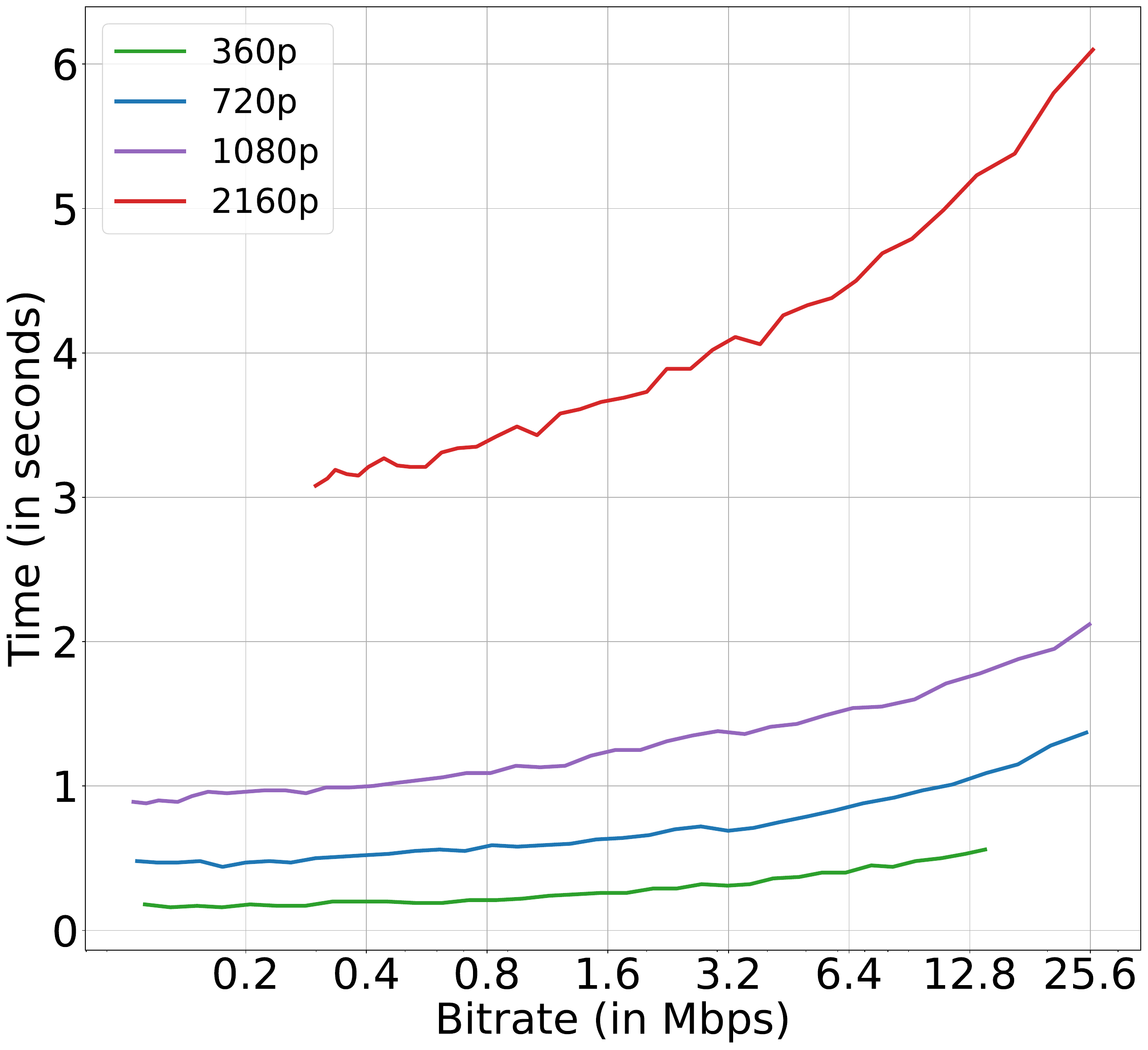}
\includegraphics[width=0.495\columnwidth]{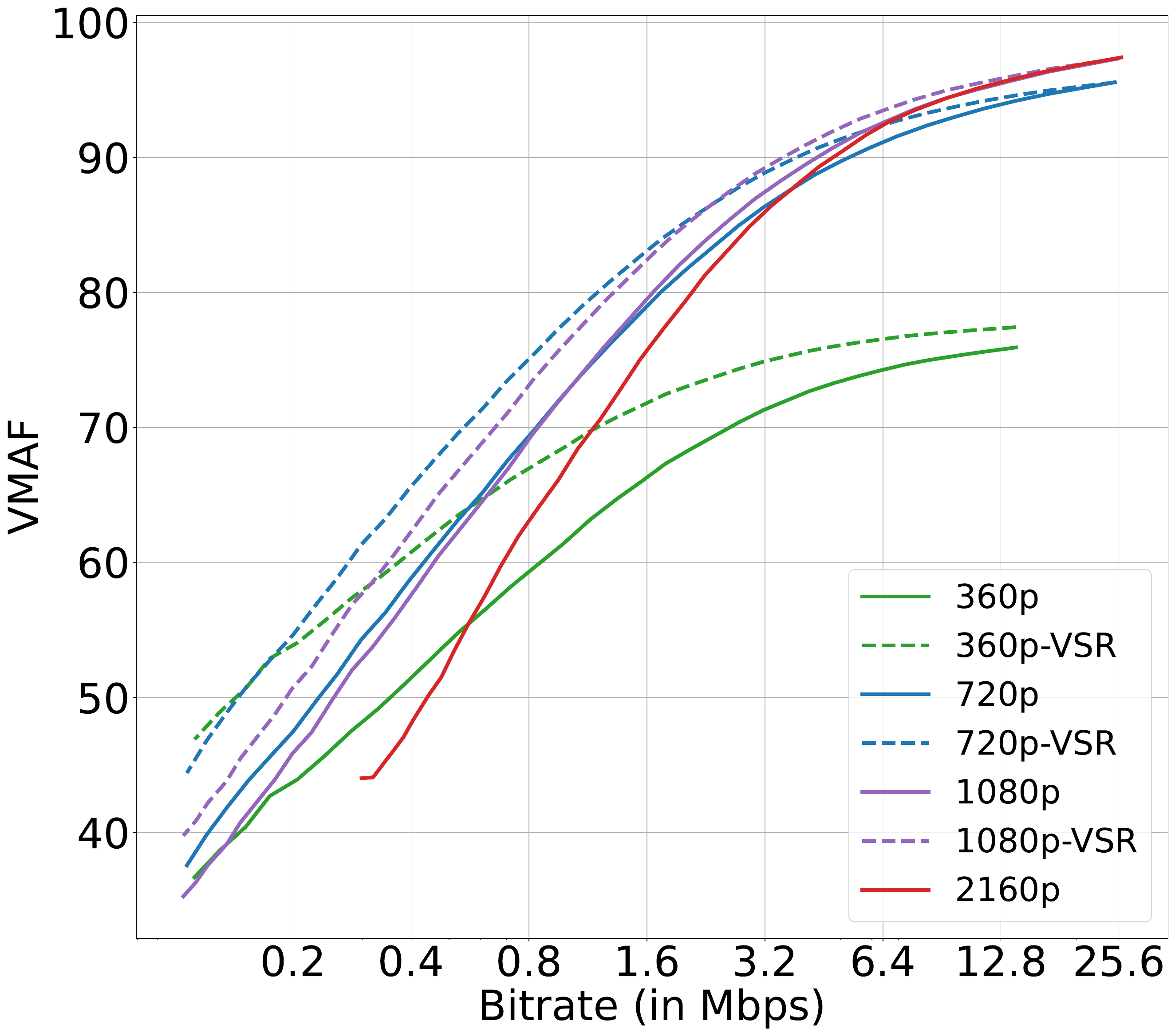}
\vspace{-0.2em}
\caption{\textit{Dolls\_s000}}
\label{fig:intro_dolls}
\end{subfigure}
\vspace{-0.6em}
\caption{Encoding results (encoding time and VMAF~\cite{VMAF}) of \textit{Characters\_s000} and \textit{Dolls\_s000} sequences of VCD dataset~\cite{VCD_ref} encoded at various resolutions, with and without client-side VSR using EDSR~\cite{edsr_ref}.}
\vspace{-1.5em}
\label{fig:vsr_intro_fig}
\end{figure}

In recent years, client devices have increased processing capability owing to the increasing RAM capacity, CPU power, and, more importantly, the introduction of powerful GPUs. Hence, it is possible to execute deep neural network (DNN)-based video super-resolution (VSR) approaches in the client devices with generic graphics processing unit (GPU) to enhance low-resolution bitstreams into high-quality streams, improving viewer retention and satisfaction~\cite{deep_vsr_survey}. Some of the popular VSR models include \texttt{FSRCNN}~\cite{fsrcnn_ref}, \texttt{ESPCN}~\cite{espcn_ref}, \texttt{EDSR}~\cite{edsr_ref}, \texttt{EVSRNet}~\cite{evsrnet_ref}, \texttt{CARN}~\cite{carn_ref}, \texttt{SRGAN}~\cite{srgan_ref}, and \texttt{RBPN}~\cite{rbpn_ref}. Even when the encoding resolution is reduced to control the encoding latency, client-side VSR can reduce the associated quality drop. As shown in Fig.~\ref{fig:vsr_intro_fig}, using \texttt{EDSR}-based client-side VSR for 360p, 720p, and 1080p representations improves visual quality (in terms of VMAF~\cite{VMAF}) at lower bitrates.  Moreover, client-side VSR helps optimize bitrate allocation, which lowers the data transmission costs of streaming. It complements traditional server-side adaptive streaming techniques and contributes to a more energy-efficient and customized video delivery strategy, ultimately improving the overall viewing experience for the audience.

Various related works in the literature explored estimating the encoding resolution to yield maximum perceptual quality~\cite{gnostic,netflix_paper,res_pred_ref1,faust_ref,jtps_ref}. However, as observed in comparing the rate-distortion curves in Fig.~\ref{fig:vsr_intro_fig}, using VSR approaches pushes the upper-quality boundary of the convex-hull~\cite{netflix_paper}, especially at lower bitrates. It is also observed that the cross-over points of the RD curves of the considered resolutions change when VSR is applied. Hence, predicting the perceptual quality of representations after VSR is mandatory to select optimized encoding resolutions for the bitrate ladder representations, which, to our knowledge, has not been investigated yet. Furthermore, it is also observed that the perceptual redundancy between representations increases with VSR. Hence, JND-aware representation elimination~\cite{cvfr_ref} can save encoding energy and streaming bandwidth costs.

\subsubsection*{Contributions} 
This paper proposes \scheme, a VSR-based encoding solution for green online video streaming.  It includes an encoding latency-aware dynamic resolution encoding scheme for online adaptive streaming applications where the clients are assumed to have GPU capability for real-time VSR. To the best of our knowledge, \scheme presents the pioneering effort in estimating optimized encoding resolution for each target bitrate, which yields the maximum perceptual quality after client-side VSR techniques, constrained to a maximum acceptable encoding latency. A random forest-based model is trained to predict the VMAF score of the representation viewed by the client. Furthermore, based on the predicted VMAF scores, perceptually redundant representations are eliminated from the bitrate ladder, saving the encoding energy and streaming costs.
\begin{figure}[t]
\centering
    \includegraphics[width=0.9\columnwidth]{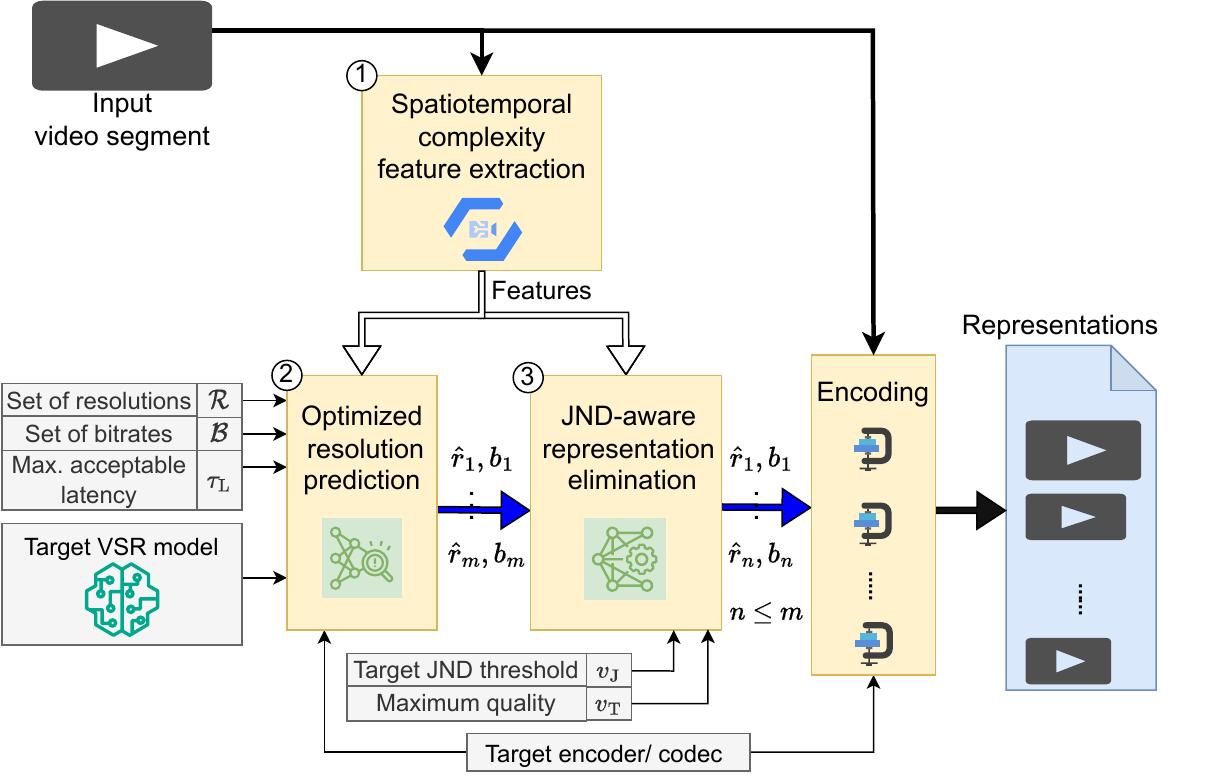}
\vspace{-0.34em}    
\caption{Encoding architecture using \scheme for green online streaming.}
\label{fig:prop_method_arch}    
\end{figure}

\section{\scheme architecture}
\label{sec:proposed_alg} 
The architecture of \scheme is shown in Fig.~\ref{fig:prop_method_arch}. \scheme receives the set of supported resolutions ($\mathcal{R}$) and set of bitrates ($\mathcal{B}$), maximum acceptable encoding latency ($\tau_{\text{L}}$), target JND ($v_{\text{J}}$), the maximum quality threshold ($v_{\text{T}}$) and the target VSR model as input. It extracts spatiotemporal complexity features of the input video segment and uses them to predict optimized encoding resolution for each target bitrate. Furthermore, it eliminates perceptually redundant representations based on the predicted quality metric (\ie VMAF in this paper) of each representation. 
\scheme comprises three phases (\cf Fig.~\ref{fig:prop_method_arch}):
\begin{enumerate}[topsep=0pt,leftmargin=*,label=\protect\circled{\arabic*}]
    \item spatiotemporal complexity feature extraction;
    \item optimized resolution prediction;
    \item JND-aware representation elimination.
\end{enumerate}
The encoding process is performed exclusively for the selected bitrate-resolution combinations for every video segment.
On the client side, VSR is applied to only low-resolution decoded representations before display.
\vspace{-0.8em}
\subsection{Spatiotemporal complexity feature extraction}
\label{sec:features}
\scheme leverages three \DCT{}energy-based features: \textit{(i)} the mean texture energy (\EY), \textit{(ii)} the mean gradient of the texture energy (\h), and \textit{(iii)} the mean brightness (\LY) extracted using \emph{video complexity analyzer} (VCA) as the spatiotemporal features of video segments~\cite{vca_ref,jtps_ref}.

\subsection{Optimized resolution estimation}
\label{sec:res_pred}

The optimized resolution selection in \scheme based on bitrate, target latency, and video complexity features is decomposed into two parts, as discussed below.

\subsubsection{Modeling} 
The perceptual quality after VSR, $v_{(r_{t},b_{t})}$ and encoding time $\tau_{(r_{t},b_{t})}$ of $t^{th}$ representation relies on video complexity features \EY, \h, \LY, encoding resolution $r_t$, and target bitrate $b_t$~\cite{csvt_ref1,cvfr_ref}:
\begin{align}
\label{eq:v_pred}
    v_{\left(r_{t},b_{t}\right)} &= f_{\text{V}}\left(E_{\text{Y}}, h, L_{\text{Y}}, r_{t}, b_{t}\right);\\
\label{eq:s_pred}
    \tau_{\left(r_{t},b_{t}\right)} &= f_{\tau}\left(E_{\text{Y}}, h, L_{\text{Y}}, r_{t}, b_{t}\right).
\end{align}

\subsubsection{Optimization}
\scheme optimizes the perceptual quality of encoded video segments after VSR while adhering to real-time processing constraints. It predicts the optimized resolution of the $t$\textsuperscript{th} representation to maximize the compression efficiency while maintaining the encoding time below the threshold  $\tau_{\text{L}}$. The optimization function is:
\begin{align}
   \hat{r}_{t} &= \argmax_{r \in \mathcal{R}} \hat{v}_{(r,b_{t})} & c.t. \hspace{1em} \hat{\tau}_{(r,b_{t})} \leq \tau_{\text{L}},  
\end{align}
where $\hat{v}_{(r,b_{t})}$ and $\hat{\tau}_{(r,b_{t})}$ are the predicted VMAF after VSR and encoding speed of the representation $(r,b_{t})$.

\subsection{JND-aware representation elimination}
\label{sec:jnd_elim}
In practice, empirical evidence demonstrates a striking similarity in VMAF scores across various representations, indicating perceptual redundancy within the bitrate ladder (as shown in Fig~\ref{fig:intro_char}). As a result, this redundancy signifies inefficient energy utilization during data encoding, storage, and transmission, failing to augment the quality of experience (QoE). To mitigate this perceptual redundancy, \scheme employs the notion of the \jnd (JND) threshold. The JND threshold signifies the minimum threshold at which the human eye can discern differences in quality~\cite{lin2015experimental,wang2017videoset,zhu2022framework, mcbe_ref}. \scheme uses the JND-aware representation elimination algorithm proposed in our previous work~\cite{cvfr_ref}. However, it is described in this paper (\cf Algorithm~\ref{algo:res_eliminate}) to make it self-contained. 
\setlength{\textfloatsep}{1pt}
\begin{algorithm}[t]
\caption{JND-based representation elimination.}
\footnotesize
\textbf{Input:}\\
\quad $m$~: number of representations in $\mathcal{B}$\\
\quad $\mathcal{Q}=\bigcup_{t=1}^m\left\{\left(\hat{r}_{t},b_{t},\hat{v}_{t}\right)\right\}$:representations with predicted resolution \\
\quad $\hat{v}_{t}~; 1 \leq t \leq m$: predicted VMAF\\
\quad $v_{\text{T}}$~: maximum VMAF threshold \\
\quad $v_{\text{J}}$~: average target JND \\
\textbf{Output:}  $\hat{\mathcal{Q}}=\left(\hat{r},b \right)$: set of  encoding configurations\\
    \begin{algorithmic}[1]
    \STATE $\hat{\mathcal{Q}} \gets \left\{\left(\hat{r}_{1},b_{1}\right)\right\}$ \label{alg:r_hat_set} \\
    \STATE $u \gets 1$  \label{alg:u_initialization}\\
    \IF{$\hat{v}_{1} \geq v_{\text{T}}$} \label{alg:vmaf_R1_comparison}
        \STATE \Return $\hat{\mathcal{Q}}$  \label{alg:retR}
    \ENDIF
        \STATE $t \gets 2$  \label{alg:t_initialization}\\ 
        \WHILE{$t \leq m$}  \label{alg:while_start}
            \IF{$\hat{v}_{t} - \hat{v}_{u} \geq v_{\text{J}}$} \label{alg:vmaf_comparison}
                \STATE $\hat{\mathcal{Q}} \gets \hat{\mathcal{Q}}\cup \left\{\left(\hat{r}_{t},b_{t}\right)\right\}$\\
                \STATE $u \gets t$  \label{alg:add_R_hat} \\
                \IF{$\hat{v}_{t} \geq v_{\text{T}}$}  \label{alg:vmaf_comp}
                    \STATE \Return $\hat{\mathcal{Q}}$ \label{alg:loop_retR}
                \ENDIF
            \ENDIF
            \item $t \gets t + 1$
        \ENDWHILE    \label{alg:while_end}      
    \STATE \Return $\hat{\mathcal{Q}}$
    \end{algorithmic}
\label{algo:res_eliminate}
\end{algorithm}

Assume the predicted VMAF difference between two representations is below $v_{\text{J}}$; consequently, the higher bitrate representation within this scope is removed. Moreover, if the predicted VMAF is greater than the maximum VMAF threshold value ($v_{\text{T}}$) at which the representation achieves perceptual losslessness, said representation is excluded from the bitrate ladder. In this manner, \scheme effectively reduces the overall encoding latency.

\section{Experimental results }
\label{sec:evaluation}

\subsection{Test methodology}
\label{sec:test_methodology}
We use 400 sequences (\SI{80}{\percent} of the sequences) from the Video Complexity Dataset~\cite{VCD_ref} as the training dataset, and the remaining (\ie \SI{20}{\percent}) is used as the test dataset. The experimental parameters used to evaluate \scheme are shown in Table~\ref{tab:exp_par}. We encode the sequences at 30\,fps with the fastest encoding preset supported by the x265~\cite{x265_ref} encoder, \ie \emph{ultrafast} preset, on a dual-processor server with Intel Xeon Gold 5218R (80 cores, frequency at 2.10\,GHz). VCA and x265 are run using eight CPU threads with x86 SIMD optimization~\cite{x86_simd_ref}. We classify the maximum acceptable encoding times ($\tau_{\text{L}}$) as ultra-low latency (\SI{1}{\second}), low latency (\SI{2}{\second}), moderate latency (\SI{4}{\second}), and standard latency (\SI{8}{\second}). We measure VSR results on an NVIDIA GeForce GT 710 GPU (frequency at 954\,MHz). Though EDSR, ESPCN, FSRCNN, and LapSRN models are available in OpenCV, we use FSRCNN for evaluation, as it yields the shortest throughput time. It is noteworthy that EDSR yields the highest VMAF improvements but with a throughput time of 430x compared to that of FSRCNN. 

We compare \scheme with the following state-of-the-art dynamic resolution estimation schemes:
\begin{enumerate}
    \item \textit{Default}: This scheme employs a fixed bitrate ladder, \ie a fixed set of bitrate-resolution pairs. This paper uses the HLS bitrate ladder specified in the Apple authoring specifications~\cite{HLS_ladder_ref} as the fixed bitrate ladder.
    \item \opte~\cite{jtps_ref}: This scheme predicts optimized resolution, which yields the highest VMAF for a given target bitrate.
\end{enumerate}
Other state-of-the-art methods in the literature~\cite{netflix_paper, faust_ref,gnostic, res_pred_ref1} are not considered for evaluation, as they are unsuitable for live streaming applications.

\begin{table}[t]
\caption{Experimental parameters used to evaluate \scheme.}
\centering
\resizebox{0.85\columnwidth}{!}{
\begin{tabular}{l||c|c|c|c|c|c}
\specialrule{.12em}{.05em}{.05em}
\specialrule{.12em}{.05em}{.05em}
\emph{Parameter} & \multicolumn{6}{c}{\emph{Values}}\\
\specialrule{.12em}{.05em}{.05em}
\specialrule{.12em}{.05em}{.05em}
$\mathcal{R}$ & \multicolumn{6}{c}{\{ 360, 720, 1080, 2160 \} } \\
\hline
$\mathcal{B}$ & 0.145 & 0.300 & 0.600 & 0.900 & 1.600 & 2.400 \\
              & 3.400 & 4.500 & 5.800 & 8.100 & 11.600 & 16.800 \\
\hline
 $\tau_{\text{L}}$ & \multicolumn{6}{c}{ 1s, 2s, 4s, 8s, $\infty$ } \\
\hline
 $v_{\text{J}}$ & \multicolumn{2}{c}{2} & \multicolumn{2}{|c}{4} & \multicolumn{2}{|c}{6}\\
\hline
 $v_{\text{T}}$ & \multicolumn{2}{c}{98} & \multicolumn{2}{|c}{96} & \multicolumn{2}{|c}{94}\\
 \hline
Target encoder & \multicolumn{6}{c}{x265 (ultrafast preset)} \\
 \hline
Target VSR & \multicolumn{6}{c}{no VSR, FSRCNN} \\
\specialrule{.12em}{.05em}{.05em}
\specialrule{.12em}{.05em}{.05em}
\end{tabular}
}
\label{tab:exp_par}
\end{table}

\begin{figure*}[t]
\centering
\begin{subfigure}{0.92\columnwidth}
    \centering
    \includegraphics[width=0.49\textwidth]{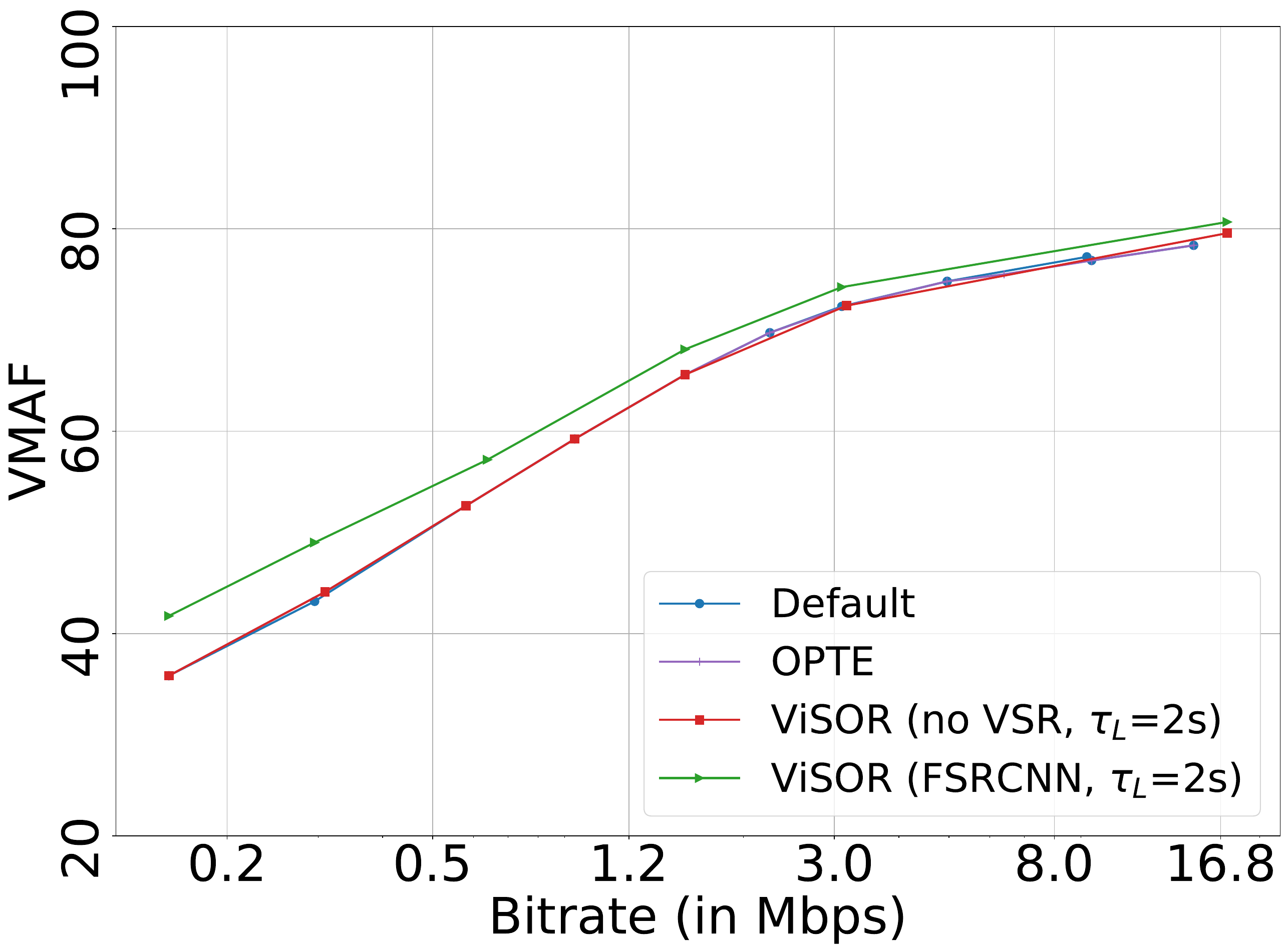}
    \includegraphics[width=0.49\textwidth]{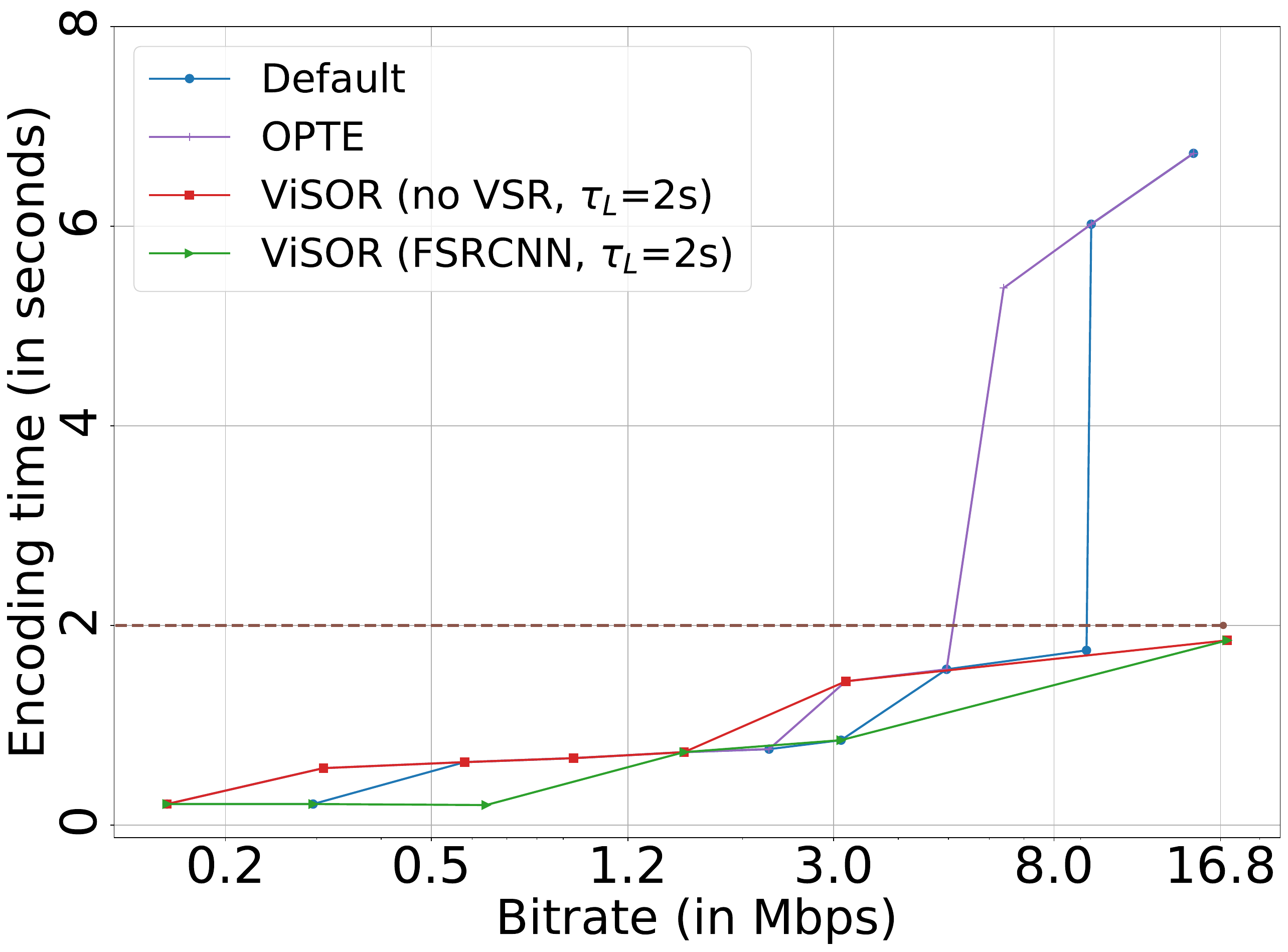}    
    \caption{\textit{Beauty\_s000}}
\end{subfigure}
\hspace{1em}
\begin{subfigure}{0.92\columnwidth}
    \centering
    \includegraphics[width=0.49\textwidth]{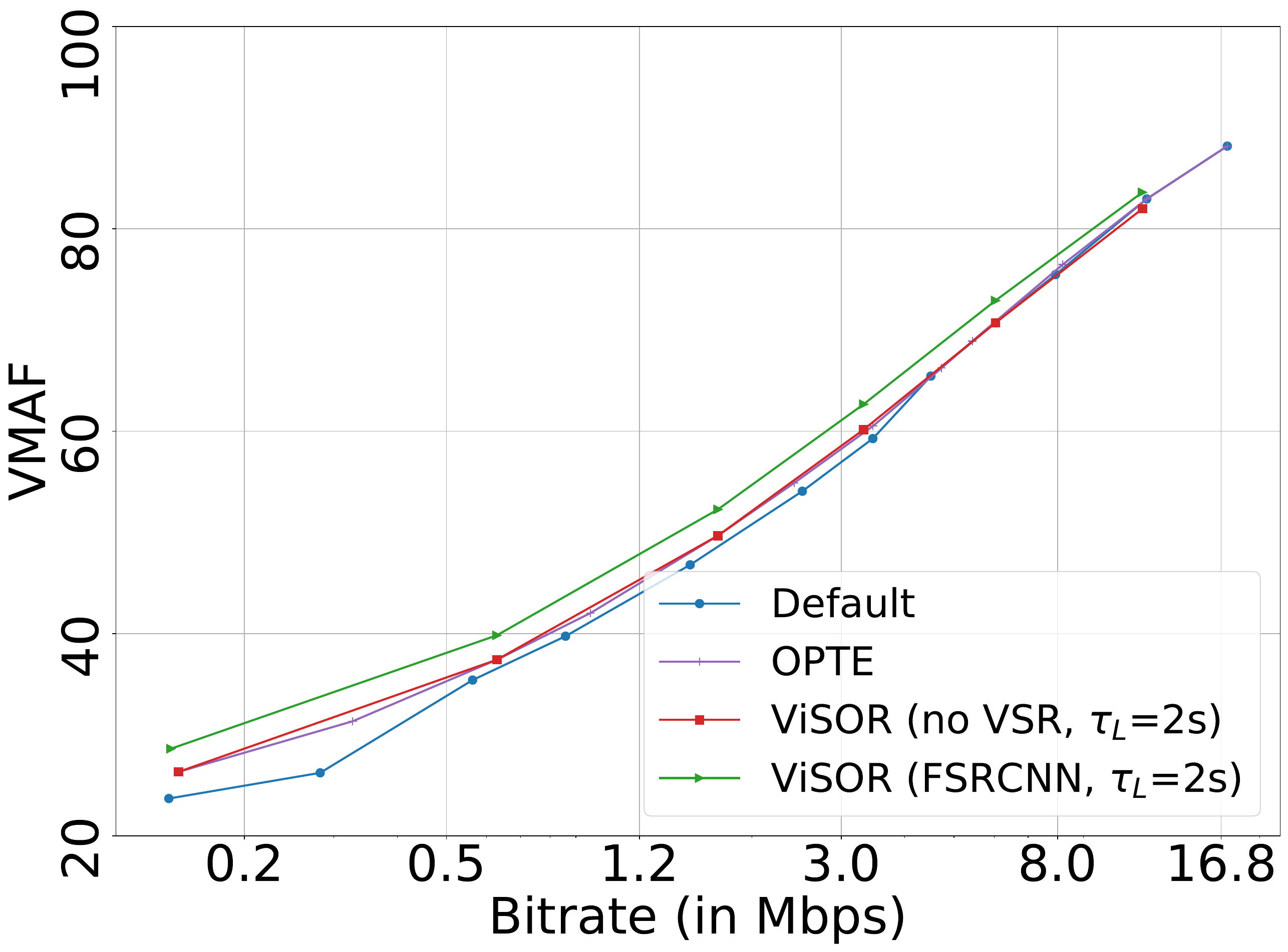}
    \includegraphics[width=0.49\textwidth]{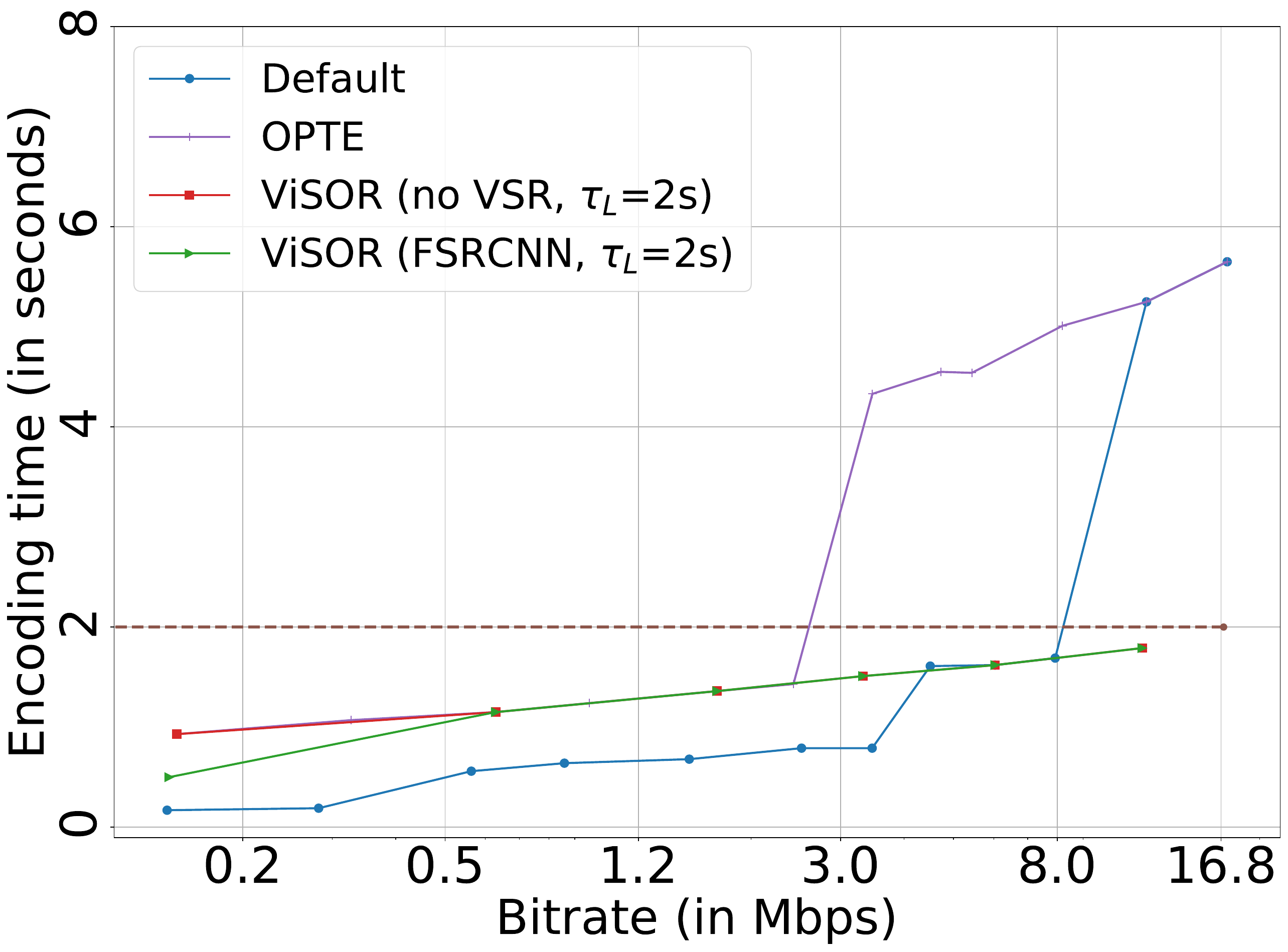}
    \caption{\textit{Park\_s000}}    
    \label{fig:bunny_rd}
\end{subfigure}
\vfill
\begin{subfigure}{0.92\columnwidth}
    \centering
    \includegraphics[width=0.49\textwidth]{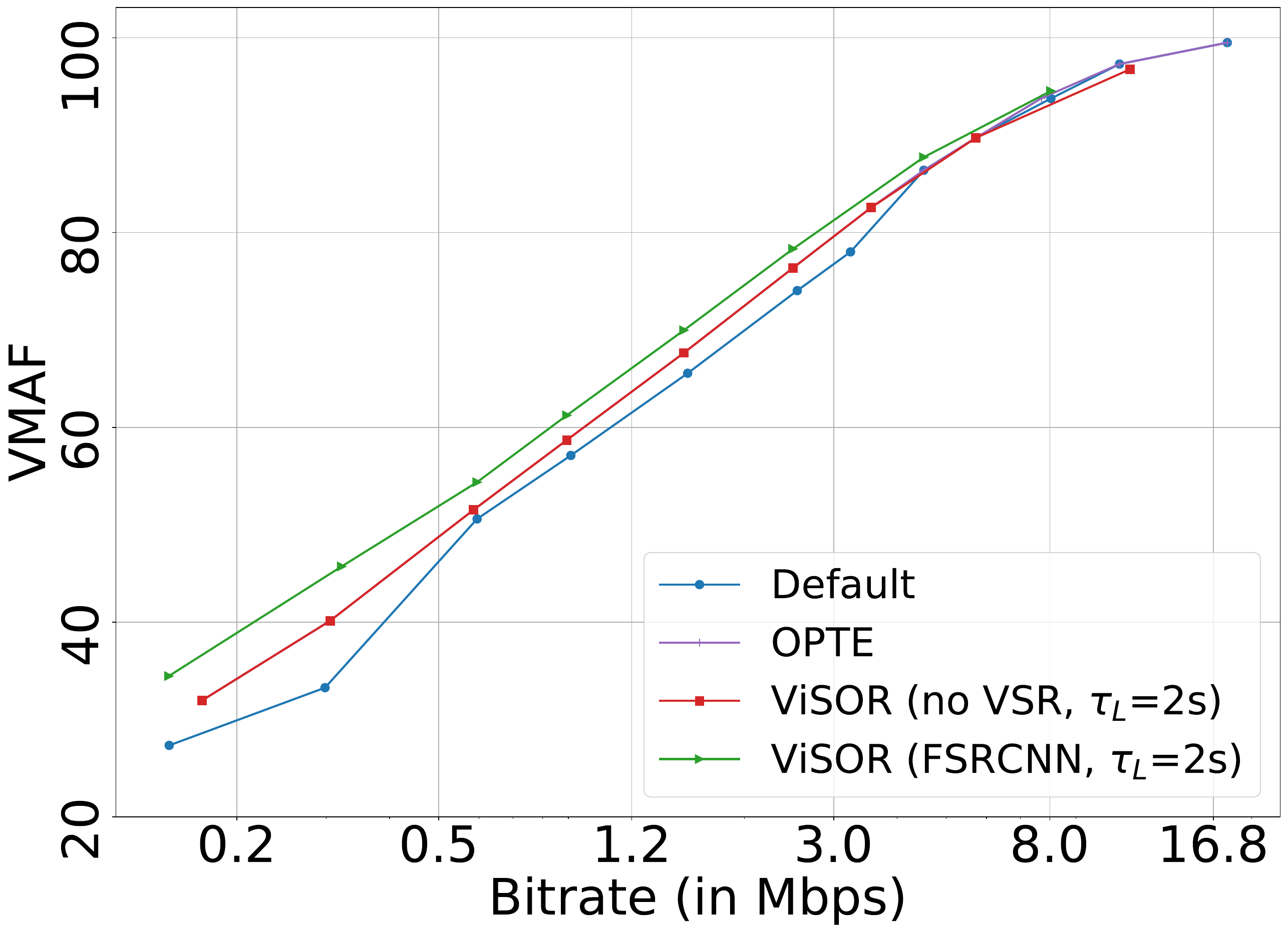}
    \includegraphics[width=0.49\textwidth]{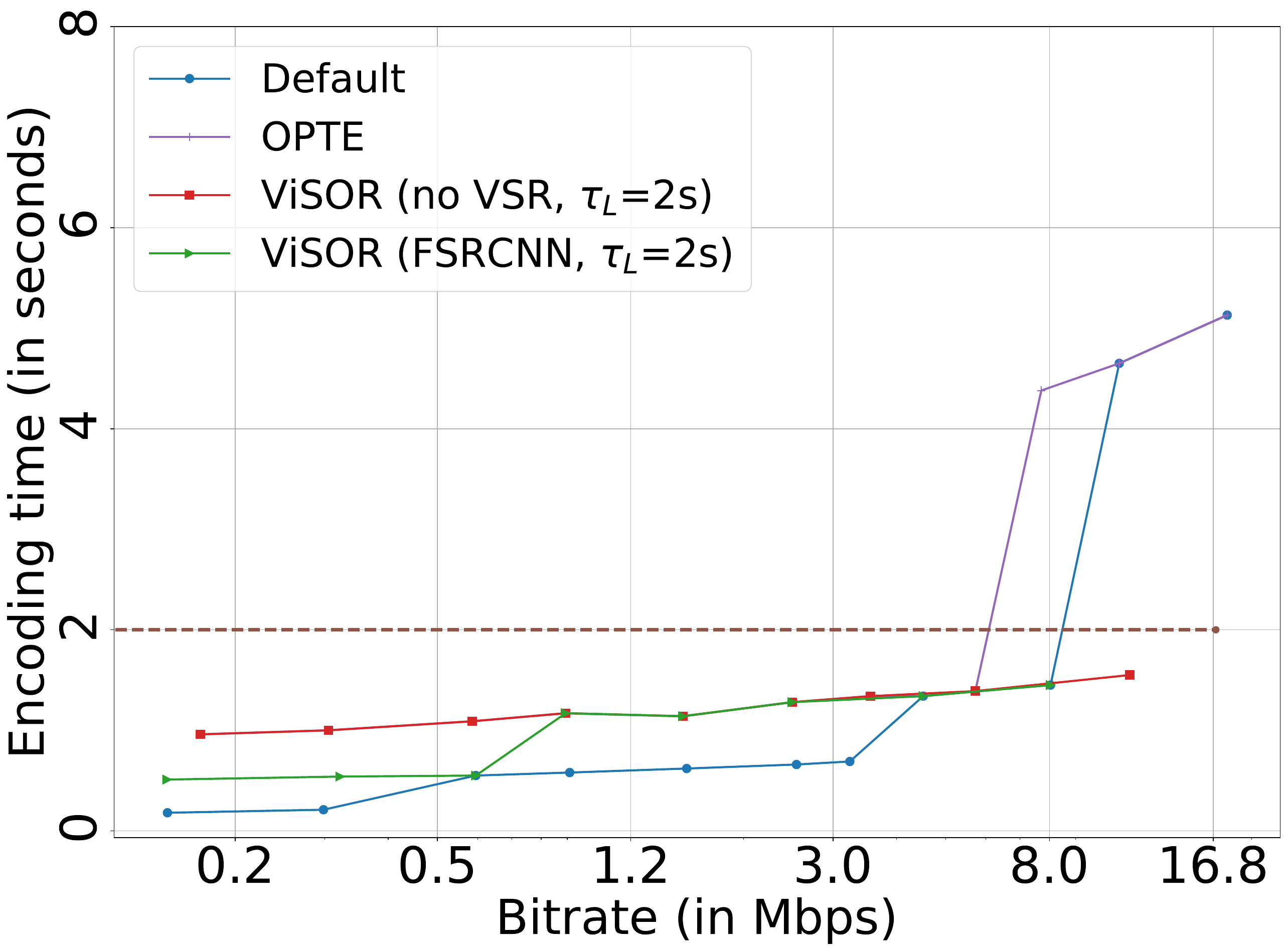}    
    \caption{\textit{RaceNight\_s000}}
\end{subfigure}
\hspace{1em}
\begin{subfigure}{0.92\columnwidth}
    \centering
    \includegraphics[width=0.49\textwidth]{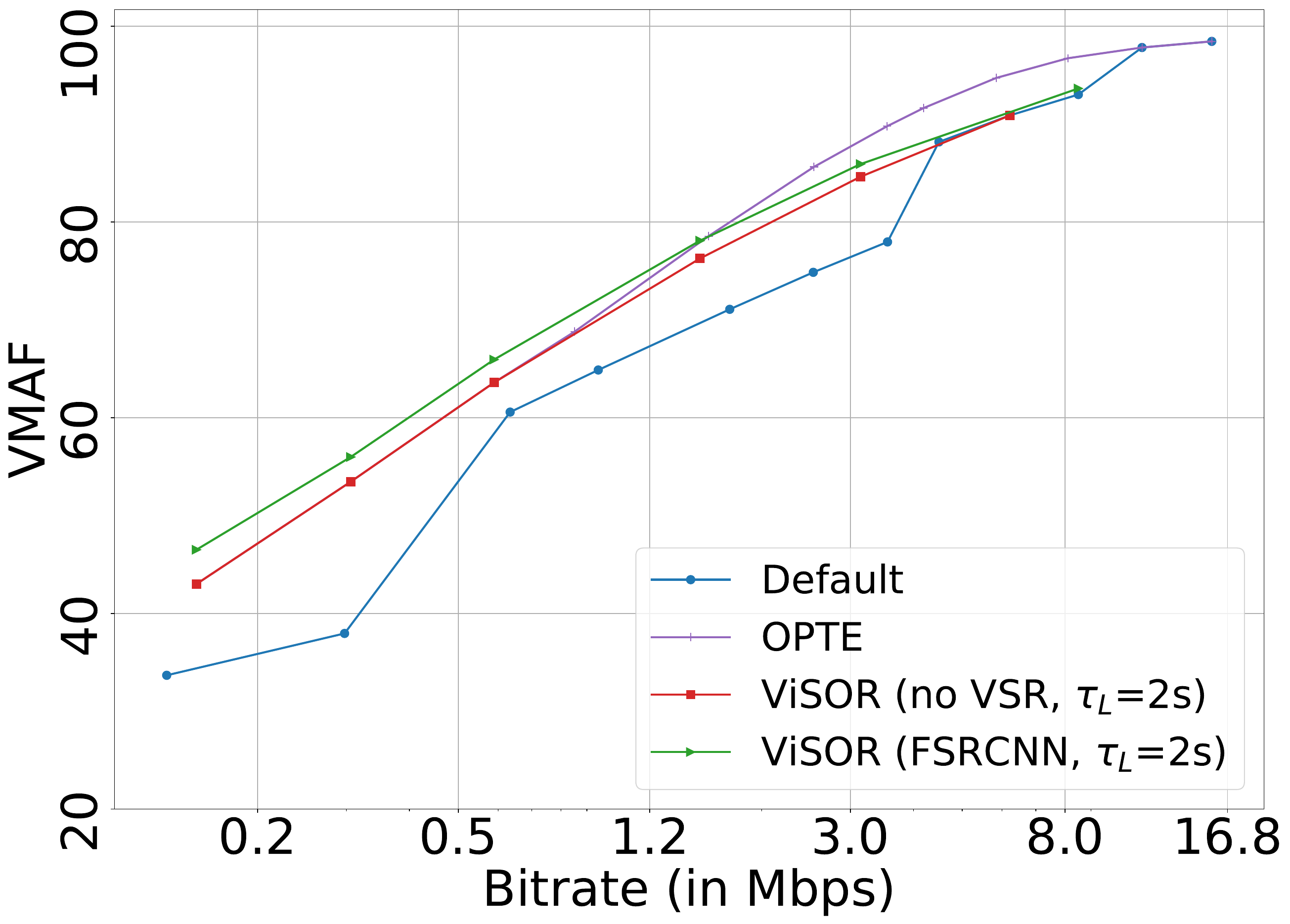}
    \includegraphics[width=0.49\textwidth]{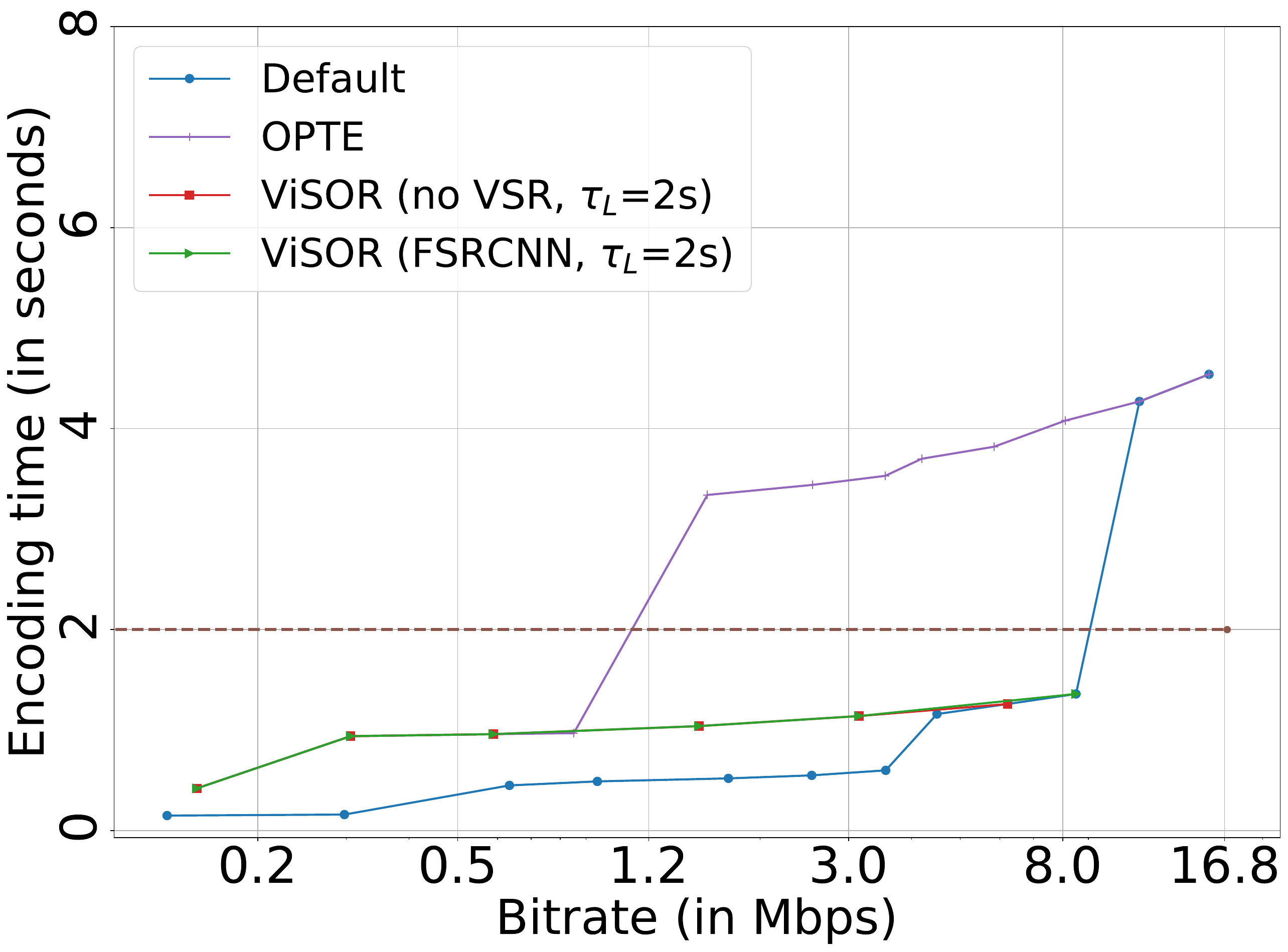}    
    \caption{\textit{Wood\_s000}}
\end{subfigure}
\vspace{-0.5em}
\caption{Rate-distortion (RD) curves and encoding times of the representative video sequences (segments) using \textit{Default} encoding (blue line), \opte (purple line), \scheme without VSR (red line), and \scheme with \texttt{FSRCNN}-based VSR (green line) for $\tau_{\text{L}}$=\SI{2}{\second}.}
\vspace{-0.5em}
\label{fig:rd_res}
\end{figure*}
\begin{table*}[!t]
\caption{Average results of \scheme compared to \textit{Default} encoding~\cite{HLS_ladder_ref}.}
\centering
\resizebox{0.995\linewidth}{!}{
\begin{tabular}{l|c|c||c|c|c|c|c|c|c||c|c|c|c|c|c|c}
\specialrule{.12em}{.05em}{.05em}
\specialrule{.12em}{.05em}{.05em}
\multicolumn{3}{c||}{Parameters} & \multicolumn{7}{c||}{Without VSR} & \multicolumn{7}{c}{With \texttt{FSRCNN}-based VSR}\\
\specialrule{.12em}{.05em}{.05em}
Remark & $\tau_{\text{L}}$ & $v_{\text{J}}$ & \BDRP & \BDRV & BD-PSNR & BD-VMAF & $\Delta E$ & $\Delta S$ & $\overline{\tau_{\text{E}}}$ & \BDRP & \BDRV & BD-PSNR & BD-VMAF & $\Delta E$ & $\Delta S$ & $\overline{\tau_{\text{E}}}$ \\
\specialrule{.12em}{.05em}{.05em}
\specialrule{.12em}{.05em}{.05em}
Highest quality (\opte~\cite{jtps_ref}) & $\infty$ & - & -20.49\% & -27.96\% & 1.37 dB & 6.25 & 91.41\% & 1.74\% & 2.54\,s & -28.33\% & -37.05\% & 1.57 dB & 7.73 & -15.86\% & -0.58\% & 1.11\,s \\
\specialrule{.12em}{.05em}{.05em}
\multirow{3}{*}{Ultra-low latency} & \multirow{3}{*}{\SI{1}{\second}} & 2 & -15.57\%   & -22.97\% & 0.54 dB & 4.64 & -46.47\% & -71.88\% & 0.71\,s & -24.75\% & -33.06\% & 0.84 dB & 6.55 & -50.49\% & -67.87\% & 0.66\,s \\
 &  & 4 & -14.22\%   & -21.43\% & 0.52 dB & 4.72 & -54.18\% & -80.32\% & 0.61\,s & -22.81\% & -32.14\% & 0.88 dB & 6.83 & -59.36\%  & -78.49\% & 0.54\,s \\
 &  & 6 & -10.91\%   & -19.04\% & 0.39 dB & 4.63 & -61.08\% & -85.35\% & 0.52\,s & -19.78\% & -30.30\% & 0.73 dB & 6.76 & -66.93\% & -83.96\% & 0.44\,s \\
\specialrule{.12em}{.05em}{.05em}
\multirow{3}{*}{Low latency} & \multirow{3}{*}{\SI{2}{\second}} & 2 & -15.96\% & -24.26\% & 0.64 dB & 5.81 & -31.66\% & -53.80\% & 0.77\,s & -27.19\% & -34.54\% & 0.97 dB & 6.58 & -46.25\% & -58.87\% & 0.71\,s \\
&  & 4 & -14.97\% & -23.53\% & 0.53 dB & 5.39 & -44.41\% & -69.05\% & 0.69\,s & -26.32\% & -33.84\% & 0.98 dB & 6.88 & -59.09\% & -71.70\% & 0.54\,s \\
&  & 6 & -13.46\% & -22.36\% & 0.48 dB & 4.56 & -51.67\% & -77.17\% & 0.57\,s & -24.65\% & -32.70\% & 0.93 dB & 6.81 & -68.21\% & -79.32\% & 0.42\,s \\
\specialrule{.12em}{.05em}{.05em}
\multirow{3}{*}{Moderate latency} & \multirow{3}{*}{\SI{4}{\second}} & 2 & -17.76\% & -26.93\% & 0.94 dB & 5.67 & -3.49\% & -46.49\% & 1.51\,s & -26.27\% & -34.91\% & 1.22 dB & 6.59 & -44.12\% & -50.29\% & 0.74\,s \\
&  & 4 & -18.64\% & -26.34\% & 0.85 dB & 5.83 & -15.95\% & -62.87\% & 1.12\,s & -25.09\% & -34.60\% & 1.19 dB & 6.89 & -56.59\% & -66.80\% & 0.58\,s \\
&  & 6 & -17.38\% & -26.22\% & 0.91 dB & 5.61 & -36.09\% & -73.98\% & 0.85\,s & -23.28\% & -33.28\% & 1.17 dB & 6.82 & -65.28\% & -77.93\% & 0.46\,s\\
\specialrule{.12em}{.05em}{.05em}
\multirow{3}{*}{Standard latency} & \multirow{3}{*}{\SI{8}{\second}} & 2 & -19.76\% & -27.87\% & 1.13 dB & 6.15 & 22.32\% & -45.02\% & 1.62\,s & -26.81\% & -35.05\% & 1.36 dB & 7.25 & -43.64\% & -49.14\% & 0.75\,s \\
&  & 4 & -19.35\% & -27.39\% & 1.22 dB & 6.25 & -8.73\% & -61.80\% & 1.21\,s & -25.86\% & -34.82\% & 1.27 dB & 7.47 & -56.11\% & -65.64\% & 0.58\,s \\
&  & 6 & -18.37\% & -26.19\% & 1.17 dB & 6.03 & -30.22\% & -72.20\% & 0.93\,s & -24.15\% & -33.54\% & 1.21 dB & 7.19 & -64.80\% & -74.58\% & 0.47\,s \\
\specialrule{.12em}{.05em}{.05em}
\specialrule{.12em}{.05em}{.05em}
\end{tabular}}
\vspace{-0.96em}
\label{tab:res_cons}
\end{table*}

\subsection{Prediction analysis}
The average mean absolute error (MAE) of VMAF and encoding time prediction models for all resolutions are 4.47 and \SI{0.22}{\second}, \vig{while the average standard deviation are 3.43 and \SI{0.47}{\second}, respectively,} which is acceptable for online streaming applications. 
We extract spatiotemporal features (\cf Section~\ref{sec:features}) at a rate of 352\,fps. The overall inference time (including the feature extraction time, VMAF prediction time, and inference time) for a \SI{4}{\second} video segment of 2160p resolution is \SI{0.37}{\second}. However, since the feature extraction and encoding are carried out as concurrent processes, the additional latency introduced by \scheme is negligible. 

\subsection{Rate-distortion analysis}
Fig.~\ref{fig:rd_res} shows the rate-distortion (RD) curves and the encoding resolution selected for the target bitrates specified in $\mathcal{B}$ of selected video sequences (segments) encoded using \scheme. At lower bitrates, \scheme reduces the compression artifacts, especially the blocking artifacts and ringing effects, improving visual quality. As the compressed video quality improves, the incremental quality gain from VSR becomes less significant. Moreover, lower resolutions are chosen when the target maximum latency is stringent since lower resolutions yield lower encoding latency. Similarly, lower resolutions are selected for lower target bitrates. Client-side VSR further enables the selection of lower encoding resolutions, which can be efficiently upscaled to higher quality without substantially compromising encoding latency or bitrate constraints. 

Table~\ref{tab:res_cons} shows the Bjøntegaard Delta values~\cite{DCC_BJDelta} of \scheme compared to \emph{Default} encoding. It is observed that the BD-rates in terms of PSNR (\BDRP) and VMAF (\BDRV) increase, while \mbox{BD-PSNR} and \mbox{BD-VMAF} values decrease as the encoding time limit decreases. 
\vspace{-0.5em}
\subsection{Encoding latency and energy consumption analysis}
Since the encodings are assumed to be carried out concurrently, the total encoding time for each segment is determined to be the highest encoding time yielded among the representations~\cite{emes_ref}. Table~\ref{tab:res_cons} shows the average encoding time for each segment ($\overline{\tau_{\text{E}}}$) using the considered encoding schemes, and relative energy consumption difference compared to \textit{Default} encoding ($\Delta E$). The \textit{CodeCarbon} tool~\cite{codecarbon_ref} is used to calculate the encoding energy. Notably, we observe a linear relationship between the encoding time and the encoding energy of every representation.

We observe that \opte yields the highest encoding time. The encoding time for a video segment decreases as $\tau_{\text{L}}$ is lowered. This reduction can be attributed to the adoption of lower resolutions as $\tau_{\text{L}}$ decreases. Furthermore, encoding time and energy consumption decreases as $v_{\text{J}}$ increases. This is due to eliminating more representations with an increase in $v_{\text{J}}$. \vig{Notably, in the current setup, energy savings on the server side are offset by greater losses on the client side (using VSR). However, it is expected that VSR approaches will evolve to use lower computational power.}

\subsection{Storage consumption analysis}
Higher \vig{absolute} relative storage difference ($\Delta S$) translates to reduced storage requirements and lower delivery costs, which can be significant for large-scale streaming platforms. Lower cumulative bitrates place less strain on the network infrastructure, reducing the risk of network congestion~\cite{farahani2022hybrid}. When the maximum acceptable encoding latency constraint is lowered, high bitrate representations, which typically require more computational resources and time to encode, are often sacrificed due to the impracticality of meeting the time constraint while maintaining high bitrates. Hence, we observe a decrease in storage consumption when $\tau_{\text{L}}$ decreases (\cf Table~\ref{tab:res_cons}). Representations upscaled using VSR often exhibit improved perceptual quality, reducing the need for high-resolution encodings at high bitrates. Consequently, bitstreams with lower bitrates can be selected to represent the same quality, which reduces the storage footprint. Hence, storage consumption decreases when using VSR. 

\section{Conclusions}
\label{sec:concl}
This paper proposed \scheme, an online energy-efficient VSR-based bitrate ladder estimation scheme for adaptive streaming applications, where optimized resolution is predicted for a given target bitrate and maximum acceptable encoding latency. \scheme includes an algorithm to determine an optimized encoding bitrate ladder, where redundant representations are eliminated based on the JND threshold. \scheme, on average, yields a bitrate reduction of \SI{24.65}{\percent} and \SI{32.70}{\percent} to maintain the same PSNR and VMAF, respectively, compared to the state-of-the-art HLS bitrate ladder encoding for an HEVC streaming session with the client using an \texttt{FSRCNN} super-resolution model, considering a JND of six VMAF points, and a maximum acceptable encoding latency of \SI{2}{\second}. This is accompanied by encoding energy and storage consumption savings of \SI{68.21}{\percent} and \SI{79.32}{\percent}, respectively. 

Future iterations of \scheme are poised to yield even more advantages with emerging codecs such as \VVC~(VVC)~\cite{vvc_ref}, given that the selection of lower resolutions implies significantly reduced computational complexity. 
In the future, \scheme can generate customized bitrate ladders optimized for various client device types (with and without GPU) according to their ability to perform VSR. This aligns with the principle of common media client data (CMCD)~\cite{cmcd_ref} enriching users' quality of experience. 
\balance
\bibliographystyle{IEEEtran}
\bibliography{references.bib}
\balance
\end{document}